\documentclass[usenatbib,twocolumn]{aastex61}
\usepackage[table]{}
\usepackage{wasysym}

\def\arcm{\hbox{$^\prime$}}
\usepackage[table]{}

\hypersetup{linkcolor=black,citecolor=black,filecolor=black,urlcolor=black}

\received{June 02, 2022}
\revised{31 July, 2022}
\accepted{September 09, 2022}
\accepted{to PASP}

\shorttitle{SPECULOOS Northern Observatory}
\shortauthors{Burdanov et al.}

\begin{document}

\title{\textbf{SPECULOOS Northern Observatory:\\searching for red worlds in the northern skies}}

\correspondingauthor{Artem Y. Burdanov (burdanov@mit.edu)}

\author[0000-0001-9892-2406]{Artem Y. Burdanov}
\affiliation{Department of Earth, Atmospheric and Planetary Science, Massachusetts Institute of Technology, 77 Massachusetts Avenue, Cambridge, MA 02139, USA}

\author{Julien de Wit}
\affiliation{Department of Earth, Atmospheric and Planetary Science, Massachusetts Institute of Technology, 77 Massachusetts Avenue, Cambridge, MA 02139, USA}

\author{Micha\"{e}l~Gillon}
\affiliation{Astrobiology Research Unit, Universit\'e de Li\`ege, All\'ee du 6 Ao\^ut 19C, B-4000 Li\`ege, Belgium}

\author{Rafael Rebolo}
\affiliation{Instituto de Astrof\'isica de Canarias, V\'ia L\'actea s/n, 38205 La Laguna, Tenerife, Spain}
\affiliation{Departamento de Astrof\'isica, Universidad de La Laguna, 38206 La Laguna, Tenerife, Spain}

\author{Daniel Sebastian}
\affiliation{School of Physics \& Astronomy, University of Birmingham, Edgbaston, Birmingham B15 2TT, United Kingdom}

\author{Roi Alonso}
\affiliation{Instituto de Astrof\'isica de Canarias, V\'ia L\'actea s/n, 38205 La Laguna, Tenerife, Spain}
\affiliation{Departamento de Astrof\'isica, Universidad de La Laguna, 38206 La Laguna, Tenerife, Spain}

\author{Sandrine Sohy}
\affiliation{Space Sciences, Technologies and Astrophysics Research (STAR) Institute, Universit\'e de Li\`ege, All\'ee du 6 Ao\^ut 19C, B\^at. B5C, 4000 Li\`ege, Belgium}

\author{Prajwal Niraula}
\affiliation{Department of Earth, Atmospheric and Planetary Science, Massachusetts Institute of Technology, 77 Massachusetts Avenue, Cambridge, MA 02139, USA}

\author{Lionel Garcia}
\affiliation{Astrobiology Research Unit, Universit\'e de Li\`ege, All\'ee du 6 Ao\^ut 19C, B-4000 Li\`ege, Belgium}

\author{Khalid Barkaoui}
\affiliation{Astrobiology Research Unit, Universit\'e de Li\`ege, All\'ee du 6 Ao\^ut 19C, B-4000 Li\`ege, Belgium}
\affiliation{Department of Earth, Atmospheric and Planetary Science, Massachusetts Institute of Technology, 77 Massachusetts Avenue, Cambridge, MA 02139, USA}
\affiliation{Instituto de Astrof\'{i}sica de Canarias (IAC), 38205 La Laguna, Tenerife, Spain} 

\author{Patricia Chinchilla}
\affiliation{Astrobiology Research Unit, Universit\'e de Li\`ege, All\'ee du 6 Ao\^ut 19C, B-4000 Li\`ege, Belgium}
\affiliation{Instituto de Astrof\'isica de Canarias, V\'ia L\'actea s/n, 38205 La Laguna, Tenerife, Spain}

\author{Elsa Ducrot}
\affiliation{Paris Region Fellow, Marie Sklodowska-Curie Action}
\affiliation{AIM, CEA, CNRS, Universit\'e Paris-Saclay, Universit\'e de Paris, F-91191 Gif-sur-Yvette, France}

\author{Catriona A. Murray}
\affiliation{Department of Astrophysical and Planetary Sciences, University of Colorado Boulder, Boulder, CO, 80309, USA}

\author{Peter P. Pedersen}
\affiliation{Cavendish Laboratory, JJ Thomson Avenue, Cambridge CB3 0HE, UK}

\author{Emmanu\"el Jehin}
\affiliation{Space Sciences, Technologies and Astrophysics Research (STAR) Institute, Universit\'e de Li\`ege, All\'ee du 6 Ao\^ut 19C, B\^at. B5C, 4000 Li\`ege, Belgium}

\author{James McCormac}
+\affiliation{Dept. of Physics, University of Warwick, Gibbet Hill Road, Coventry CV4 7AL, UK}

\author{Sebasti\'an Z\'u\~niga-Fern\'andez}
\affiliation{Astrobiology Research Unit, Universit\'e de Li\`ege, All\'ee du 6 Ao\^ut 19C, B-4000 Li\`ege, Belgium}




\begin{abstract}
SPECULOOS is a ground-based transit survey consisting of six identical 1-m robotic telescopes. The immediate goal of the project is to detect temperate terrestrial planets transiting nearby ultracool dwarfs (late M-dwarf stars and brown dwarfs), which could be amenable for atmospheric research with the next generation of telescopes. Here, we report the developments of the northern counterpart of the project -- SPECULOOS Northern Observatory, and present its performance during the first three years of operations from mid-2019 to mid-2022. Currently, the observatory consists of one telescope, which is named Artemis. The Artemis telescope demonstrates remarkable photometric precision, allowing it to be ready to detect new transiting terrestrial exoplanets around ultracool dwarfs. Over the period of the first three years after the installation, we observed 96 objects from the SPECULOOS target list for 6000\,hours with a typical photometric precision of $0.5\%$, and reaching a precision of $0.2\%$ for relatively bright non-variable targets with a typical exposure time of 25\,sec. Our weather downtime (clouds, high wind speed, high humidity, precipitation and/or high concentration of dust particles in the air) over the period of three years was 30\% of overall night time. Our actual downtime is 40\% because of additional time loss associated with technical problems.
\end{abstract}

\keywords{telescopes - techniques: photometric - planets and satellites: detection}



\section{Introduction} \label{sec:intro}

The successful launch of the James Webb Space Telescope (JWST, \citealt{2006SSRv..123..485G}) and current progress in  building of the ground-based extremely large telescopes (ELTs) are opening a new era of in-depth studies of extrasolar planets \citep{2014PASP..126.1134B,2015PASP..127..311C,2016ApJ...817...17G}. Specifically, detailed atmospheric characterization of transiting temperate terrestrial exoplanets will soon be possible \citep{deWit2013,2013ApJ...764..182S,2017ApJ...850..121M}, and it will provide us with the first opportunity to detect chemical traces of life beyond our solar system \citep{Kaltenegger2009, Seager2009}. This opportunity relies on the discovery of suitable targets, i.e., habitable terrestrial planets transiting a host star bright and small enough to lead to adequate signal-to-noise ratios for detection of biosignatures through eclipse spectroscopy \citep{2016AsBio..16..465S}. In this framework, the best target for biosignatures detection would be a habitable terrestrial planet transiting one of the nearest ultracool dwarfs (UCDs), i.e., very-low-mass stars and brown dwarfs with effective temperatures lower than 2700\,K, luminosities smaller than $\mathrm{10^{-3}\,L_{\odot}}$, where $\mathrm{L_{\odot}}$ is the solar luminosity, and spectral types later than M6 \citep{2005ARA&A..43..195K, 2011ApJ...743...50C}.

One of the first surveys, which tried to explore transiting exoplanets around UCDs was conducted using the Peters Automated InfRared Imaging TELescope (PAIRITEL), which observed a small set of 13 UCDs in 2004 and 2005 for a period of 10 months \citep{2008PASP..120..860B}. This project did not report any detections, most likely because of the small number of explored targets and the small number of accumulated observing hours per target. Though not uniquely focused on UCDs, APACHE \citep{2013EPJWC..4703006S} and MEarth \citep{2015csss...18..767I} ground-based surveys have been exploring mid-to-late M-dwarfs for the presence of transiting exoplanets since 2010s. The first rocky exoplanet amenable for atmospheric research was discovered around an M4.5 dwarf star GJ1132 by MEarth in 2015 \citep{2015Natur.527..204B}.

TESS (Transiting Exoplanet Survey Satellite; \citealt{2014SPIE.9143E..20R}) is designed for detecting planets around G to mid M-dwarf stars and its transit detection sensitivity is low for M-dwarf stars later than M5 due to their faintness in TESS filter. However, TESS exoplanet transit detection is possible for the brightest late M-dwarfs (see two super-Earths discovered by TESS around M5 dwarf star LHS~3844 \citep{2019ApJ...871L..24V} and M6 dwarf star LP~791-18 \citep{2019ApJ...883L..16C}). 

Ongoing surveys, such as ExTrA \citep{Bon2015} and EDEN \citep{2020AJ....159..169G} targeting specifically late M-dwarfs, and PINES \citep{2022AJ....163..253T} targeting L- and T-type dwarfs have recently started their operations and have not reported any detections yet. To date, there is only one known system with transiting planets orbiting a UCD: it is the TRAPPIST-1 system with seven exoplanets that form a unique near-resonant chain. Remarkably, three of the TRAPPIST-1 planets are located in the habitable zone of the host star \citep{2016Natur.533..221G,2017Natur.542..456G,2017NatAs...1E.129L,2021PSJ.....2....1A}. This planetary system was discovered in the context of the TRAPPIST UCD transit survey \citep{2013EPJWC..4703001G,2020MNRAS.497.3790L}. This survey has observed 50 brightest southern UCDs for about 100\,hr each with the TRAPPIST-South telescope located at the La Silla Observatory in Chile \citep{2011Msngr.145....2J,2011EPJWC..1106002G}. Most importantly, the survey served as a prototype for a more ambitious search for exoplanets around UCDs -- SPECULOOS.

SPECULOOS stands for \textbf{S}earch for habitable \textbf{P}lanets \textbf{EC}lipsing \textbf{UL}tra-c\textbf{OO}l \textbf{S}tars and it is a ground-based transit survey led by the University of Li\`ege (Belgium), which consists of six identical 1-m robotic telescopes. The immediate goal of the project is to detect temperate terrestrial planets transiting nearby ($< 40\,\mathrm{pc}$) UCDs, which are bright enough in the near-IR to make possible the atmospheric characterization of their planets with JWST and ELTs. Upon completion of the survey, we will be able to reach the ultimate goal: determining the frequency of short-period Earth-sized planets around UCDs and further constrain planet formation theories, which currently predict UCD planets ranging from metal-rich Mercury-sized planets to volatile-rich Earth-sized planets (e.g., \citealt{2007ApJ...669..606R,2009Icar..202....1M,2020MNRAS.491.1998M}). 

SPECULOOS consists of three nodes: four telescopes are installed at the ESO Paranal Observatory (Atacama Desert, Chile) and they compose the SPECULOOS Southern Observatory (SSO), which has been operational since January 2019. The second node is SPECULOOS Northern Observatory (SNO), which is currently composed of one 1m-aperture telescope that is located at the Teide Observatory (Canary Islands, Spain) and which has been operational since June 2019. The third node is the SAINT-EX telescope (Search And characterIsatioN of Transiting EXoplanets; \citealt{2020A&A...642A..49D}). It is a 1-m telescope located at the National Astronomical Observatory of Mexico (San Pedro M\'artir, Mexico), which contributes to the SPECULOOS project by observing UCD targets for 80\% of its observation time since March 2019. For the next 10 years, these 6 telescopes aim to observe 1700 nearby UCDs brighter than $K$=12.5\,mag to search for TRAPPIST-1-like systems.

\begin{figure}[ht!]
\begin{center}
\includegraphics[height=0.5\textwidth]{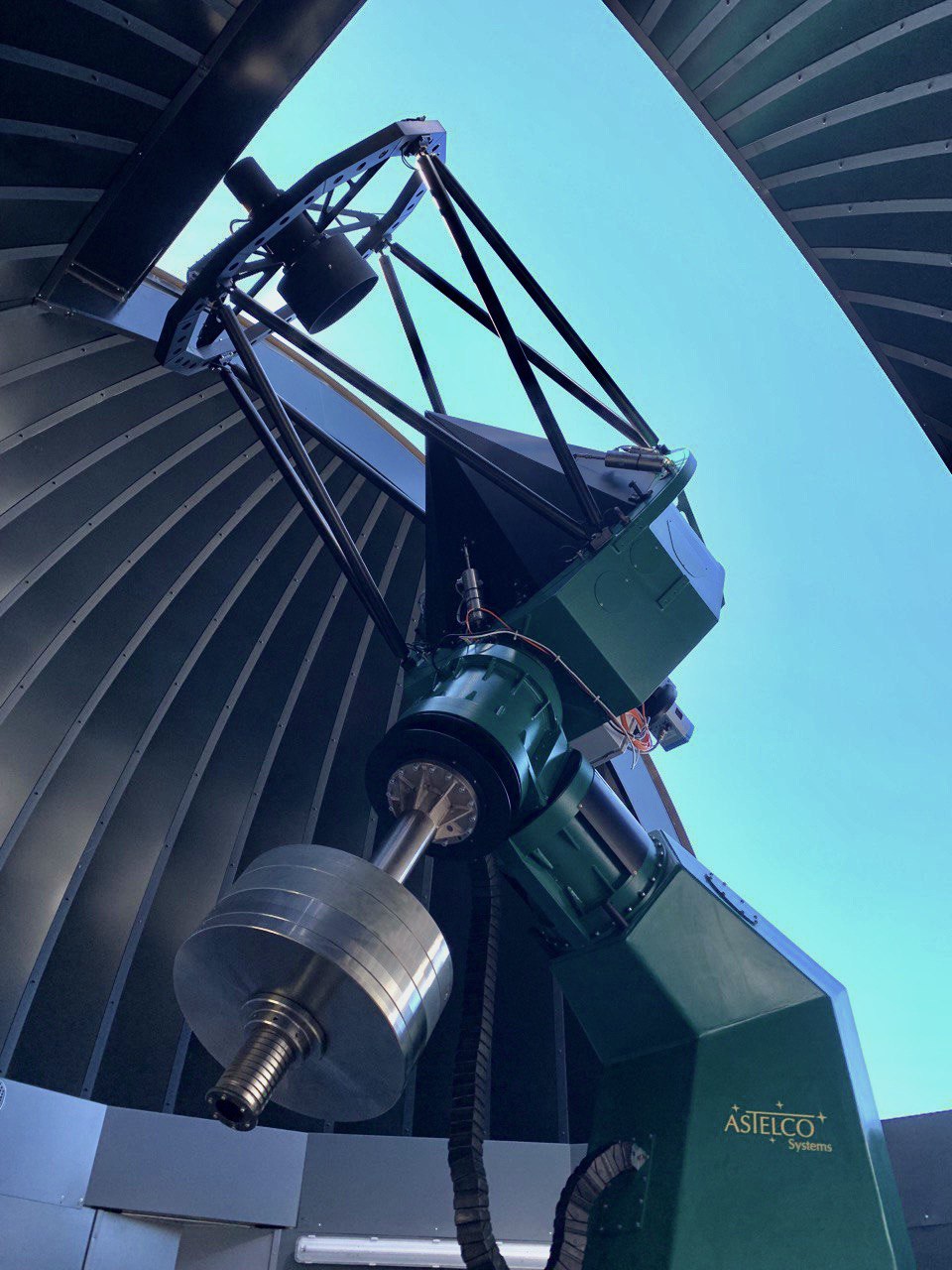}
\caption{The Artemis telescope. Note a set of lightweight black metallic petals attached to the optical tube assembly, which protect the primary mirror when the telescope is not observing.}
\end{center}
\label{fig:sno}
\end{figure}

\begin{figure*}[ht!]
\begin{center}
\includegraphics[height=5.75cm]{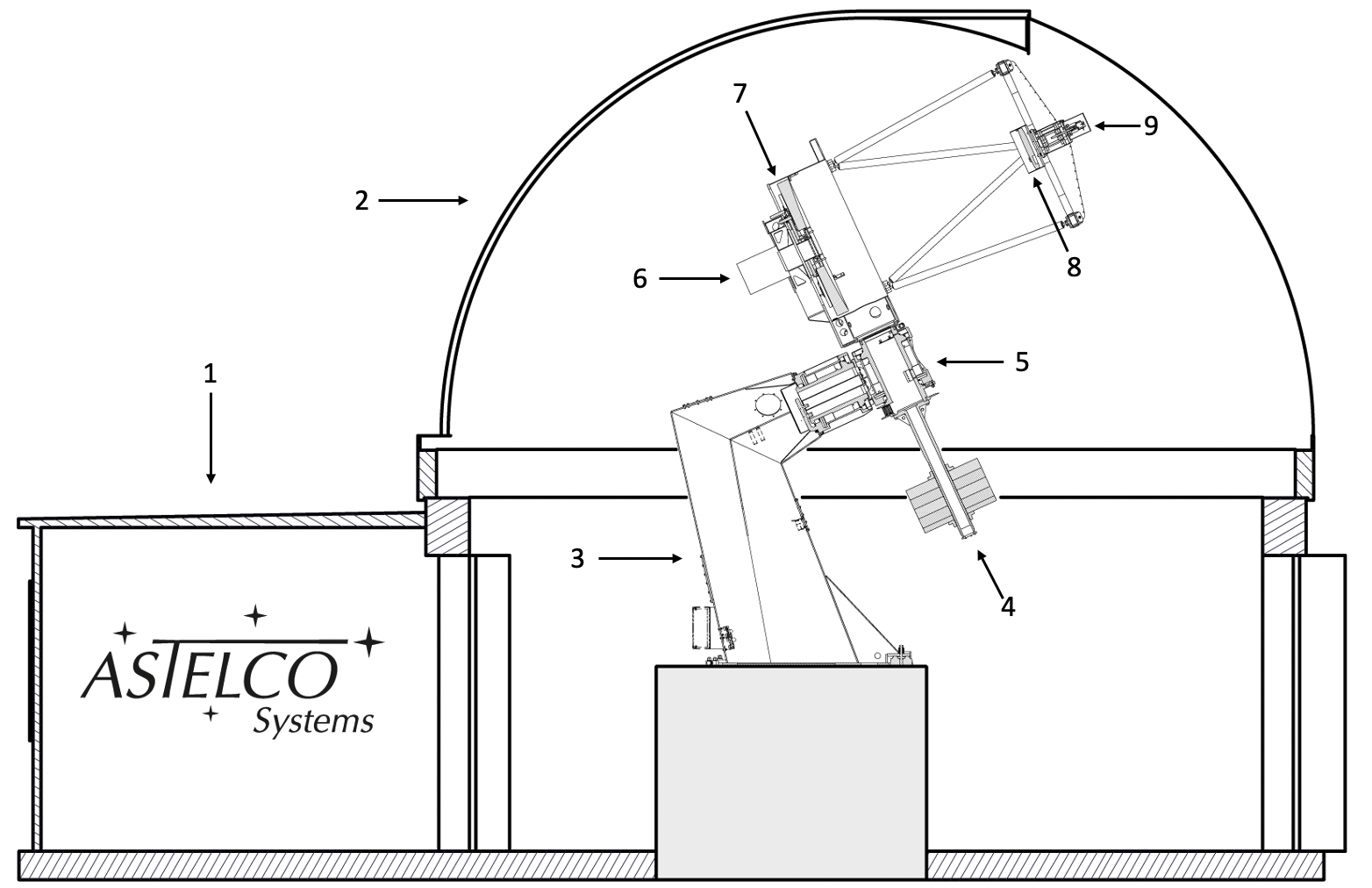}
\includegraphics[height=5.75cm]{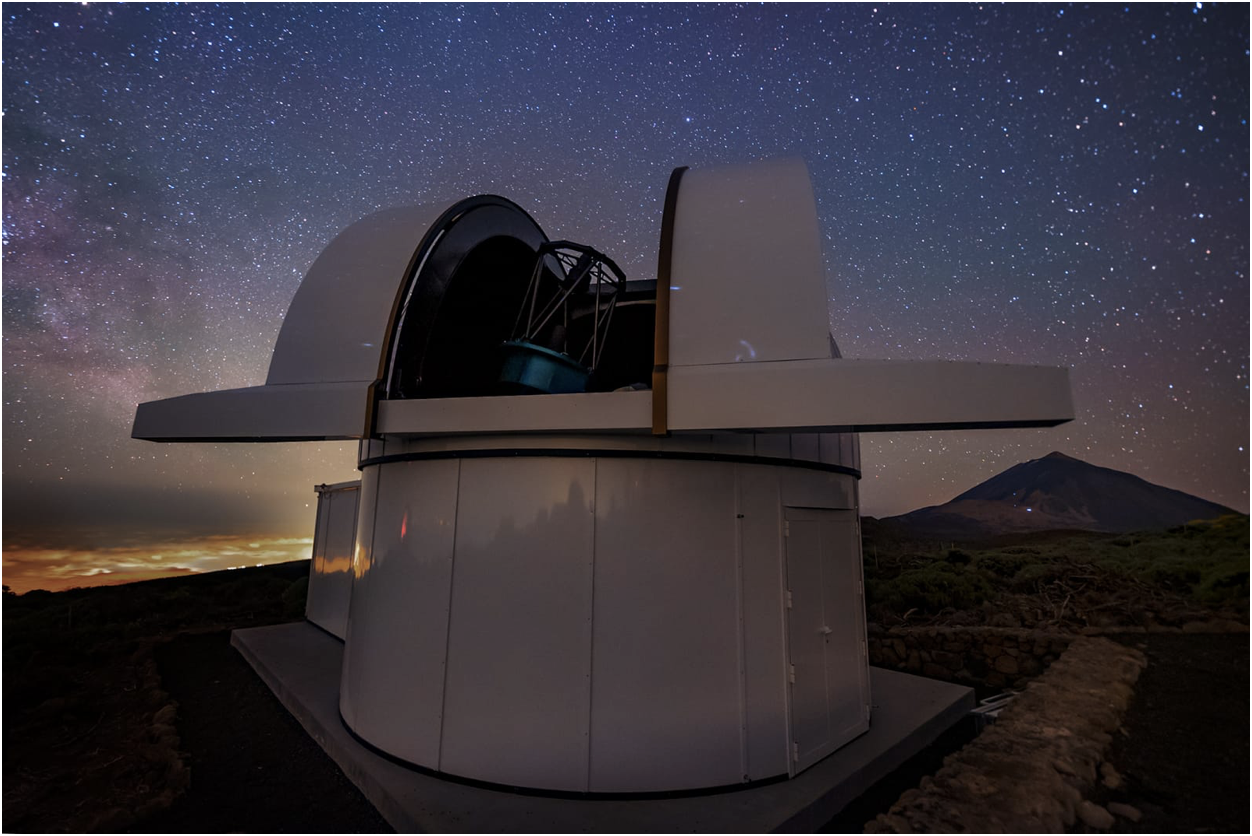}
\end{center}
\caption{Left: Technical drawing of the facility (credit: Astelco Systems): 1 -- control room, 2 -- dome, 3 -- pillar, 4 -- counterweight, 5 -- direct drive NTM-1000 mount, 6 -- CCD camera and filter wheel, 7 -- primary mirror, 8 -- secondary mirror, 9 -- focuser. Right: View of the Artemis dome with the Teide volcano in the background (credit: D.~López).}
\label{fig:sno-outside}
\end{figure*} 

\cite{2018NatAs...2..344G}, \cite{2018Msngr.174....2J}, and \cite{2018haex.bookE.130B} provide a general overview of the SPECULOOS project, while the work of \cite{2018SPIE10700E..1ID} provides technical details of the survey and evaluates preliminary photometric performance of two newly installed telescopes of SSO. Performance of all four SSO telescopes during the first year of operations and a dedicated SPECULOOS photometric data reduction pipeline are presented in \cite{2020MNRAS.495.2446M}. The survey target list and observational strategy for different SPECULOOS programs are presented in \cite{2021A&A...645A.100S}, whereas the latest development of the project as a whole can be found in the work of \cite{2020SPIE11445E..21S}. 

In this paper, we report the developments of SNO and present its performance during the first three years of operations from mid-2019 to mid-2022. The paper is organized as follows: in Section~\ref{sec:SNO_facility}, we present technical information about the observatory and its operations. Section~\ref{sec:weather} is dedicated to the discussion of statistics of observations. In Section~\ref{sec:photometry}, we describe photometric performance of the first SNO telescope and provide examples of early scientific results. We review some of the complementary science projects in Section~\ref{sec:complementary}. In Section~\ref{sec:discuss_conclusions}, we discuss the results from the first three years of operations and outline future prospects of the observatory.

\section{SPECULOOS North}\label{sec:SNO_facility}

SNO is located at the Teide Observatory on the island of Tenerife (Canary Islands, Spain). It is envisioned as a twin observatory to SSO, and it is operated by Massachusetts Institute of Technology (MIT, USA) and the University of Li\`ege (ULi\`ege, Belgium), in collaboration with the Instituto de Astrof\'isica de Canarias (IAC, Spain). Currently, SNO is composed of one telescope, which is named Artemis\footnote{Following the Greek mythology naming as in the case of SPECULOOS South telescopes (where the telescopes were named after the Galilean moons -- Europa, Io, Callisto and Ganymede), the first SPECULOOS North telescope is named after Artemis -- the goddess of wilderness, vegetation and hunting.} (see Fig.~\ref{fig:sno} and Fig.~\ref{fig:sno-outside}). It was installed in April and commissioned in June 2019. All technical information is summarized in Table~\ref{tab:general_info}.

\begin{table}[h]
\centering 
\caption{General information about the Artemis telescope.}
\begin{tabular}{ll}
\hline
\hline
Geographical coordinates & 28$^\circ$18\arcm01.44\arcm\arcm\,N,\\
& 16$^\circ$30\arcm41.04\arcm\arcm\,W,\\
& 2440\,m\\
Diameter of the primary mirror & 1\,m\\
Diameter of the secondary mirror & 0.28\,m\\ 
Focal length & 8\,m\\
Focal ration & f/8\\ 
Mount & Direct-drive German\\
& equatorial NTM-1000\\
CCD name & Andor iKon-L\\ 
& BEX2-DD-9TW\\
CCD size & $\mathrm{2K\times2K}$\\
CCD pixel size & 13.5\,$\mu$m\\
CCD pixel scale & 0.35\,$\mathrm{arcsec\,pixel}^{-1}$\\
Field of view & $12\,\times12\,\mathrm{arcmin^2}$\\
Filter wheel & FLI CFW3-10\\
Installed filters & Sloan-$g'$, -$r'$, -$i'$, -$z',$\\
& $I+z'$, blue-blocking,\\ 
& z cut\\
\hline
\end{tabular}
\label{tab:general_info}
\end{table}

Artemis is a Ritchey-Chr\'etien telescope with 1-m primary and 0.28-m secondary mirrors, and a correction lens, which together provide a focal length of 8\,m (focal ratio f/8). Both mirrors are coated with pure aluminium. The telescope was manufactured in Germany by Astelco Systems\footnote{\url{www.astelco.com}}. In contrast to SSO telescopes, Artemis has a set of lightweight metallic petals attached to the optical tube assembly (OTA; See Fig.~\ref{fig:sno}). These petals are closed during the daytime and protect the primary mirror from dust particles and various debris coming from the local flora and fauna. OTA is installed on the German equatorial mount NTM-1000 with direct drive motors, which provides precise pointing and is optimized for optimal tracking performance with no periodic errors and with no need for hardware guiding. However, a software autoguiding algorithm \texttt{DONUTS} is utilized to keep stars on the same spots of the CCD (charge-coupled device)  with sub-pixel precision to improve photometric performance \citep{2013PASP..125..548M}. Thanks to the special mount pillar, the telescope has no meridian flip and allows night-long tracking of targets (see the left panel of Fig.~\ref{fig:sno-outside}). 

The telescope is enclosed in a circular 6.2-5m classic robotic Astelco dome, with its controls integrated into the telescope’s control software (TCS). Adjacent to the circular dome building, a control room keeps the telescope's electrical cabinet and control computers. The control room of the Artemis telescope is also designed to serve as a hub of operations for planned additional SNO telescopes.

Similar to other SPECULOOS telescopes, Artemis is equipped with an Andor\footnote{\url{www.andor.oxinst.com}} iKon-L Peltier-cooled deep-depletion fringe-suppressed CCD camera (BEX2-DD-9TW). Its size is $\mathrm{2K\times2K}$ with a pixel size of 13.5~$\mu$m. The field of view is $12\,\times12\,\mathrm{arcmin^2}$ and the corresponding pixel scale is 0.35\,$\mathrm{arcsec\,pixel}^{-1}$.  It is usually operated at 1\,MHz readout mode, at -60$^\circ$C with a gain of 1\,e$^{-}$/ADU, read-out noise of 6\,e$^{-}$, and dark current of 0.1 e$^{-}$/s/pixel. The CCD detector is sensitive from near-UV to the near-IR (350-950\,nm), with a maximum quantum efficiency of 94\% at both 420 and 740\,nm. 

The CCD camera is coupled with a Finger Lake Instruments\footnote{\url{www.flicamera.com}} CFW3-10 filter wheel for ten 50-mm square filters. The following set of filters is currently installed at Artemis: Sloan-$g'$, -$r'$, -$i'$, -$z'$ filters, custom exoplanet filters $I+z'$ (transmittance $>$90\% from 750\,nm to beyond 1000\,nm) and "blue-blocking" (transmittance $>$90\% from 500\,nm to beyond 1000\,nm), and a "z cut" filter, which suppresses the effect of atmospheric water absorption (transmittance $>90\%$ from 860\,nm to 1100\,nm).  This setup is optimized to observe UCDs up to $J$=14\,mag, and to obtain their high-precision light curves ($\sim$0.1\%) with a sampling time of a few minutes. 

We note that the wide near-IR $I+z'$ filter is our standard one and it is used for observing the majority of the SPECULOOS targets. Usage of the  $I+z'$ filter allows us to minimize the effect of atmospheric extinction, which is most prominent in blue wave-range and obtain maximum photons from UCDs as their spectral energy distribution peak at near-and mid-IR wavelengths (see Fig.~7 from \citealt{2018SPIE10700E..1ID} for the $I+z'$ filter transmission curve).

\subsection{Auxiliary devices}

\begin{figure}[ht!]
\begin{center}
\includegraphics[width=0.4\textwidth]{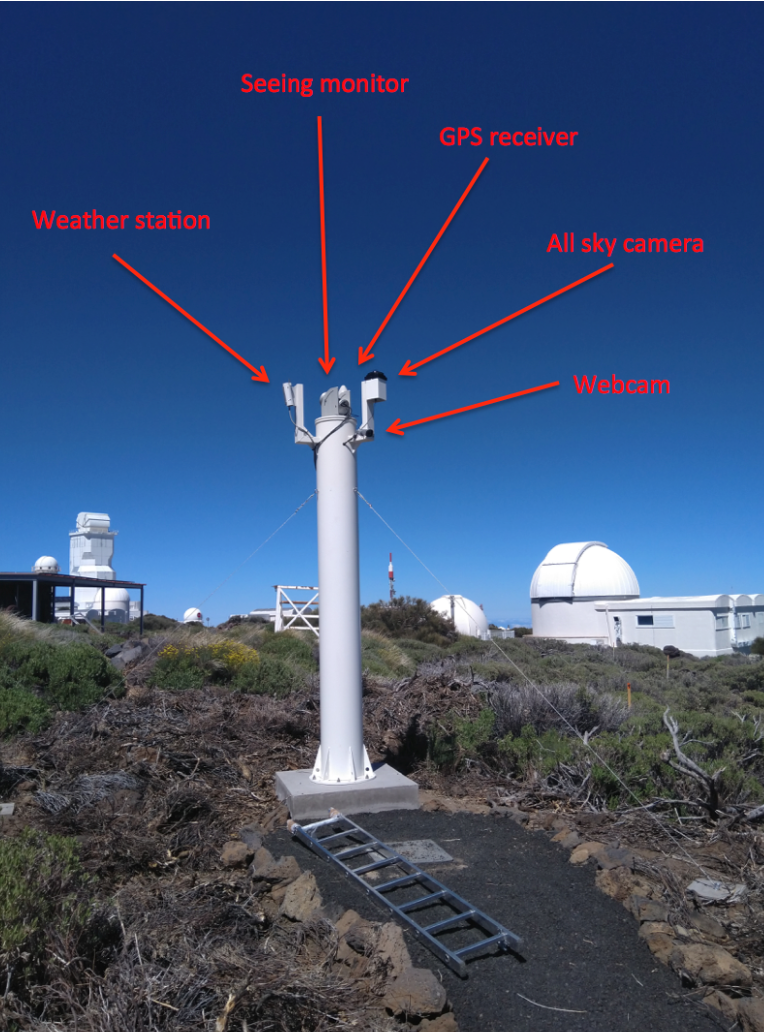}
\end{center}
\caption{The mast with auxiliary devices: weather station, seeing monitor, GPS receiver, all sky camera, webcam.}
\label{fig:mast}
\end{figure}

SNO has a broad set of auxiliary devices, which enable smooth and safe operations of the facility. Specifically, the Boltwood\footnote{\url{www.diffractionlimited.com/product/boltwood-cloud-sensor-ii/}} Cloud Sensor II is used as a weather station. It is installed on a mast near the telescope building (Fig.~\ref{fig:mast}). It measures the presence of rain droplets, proximity to a dew point, ambient temperature, cloud cover using an IR sensor, wind speed using a friction sensor, relative humidity and amount of sunlight. The Boltwood sensors trigger weather alerts, which are received by a control computer. It also features direct connection to the dome, which enables closure of the slit in case the control computer is not operational. Additionally, an independent rain sensor is installed on the telescope building. It can trigger dome closure if the Boltwood weather station is not working.

The mast also has a GPS receiver (164DHS from Meinberg Radio Clocks GmbH), which is used as a source of precise time, and the Alcor System Cyclope seeing monitor\footnote{\url{www.alcor-system.com/new/SeeingMon/Cyclope.html}}, which is used to measure seeing using the Polaris. An outside webcam and the SBIG 340 all sky camera\footnote{\url{www.diffractionlimited.com/product/all-sky-340-cameras/}} are installed on the mast as well and are used by the telescope operators to assess weather conditions.

An uninterruptable power supply (UPS) keeps running Artemis for at least 30\,min during an electrical power
cut and an emergency shutdown is triggered at the end of this period.

\subsection{Operations}\label{subsec:operations}

The Artemis telescope is controlled by the \texttt{ACP} Observatory Control Software\footnote{\url{www.acp.dc3.com}}, which provides a high level of automatization. A human telescope operator is needed only during the startup: the operator ensures the successful start of an \texttt{ACP} observing sequence before the local sunset. After that, \texttt{ACP} handles all the following operations: opening the dome after sunset, taking flat images using twilight sky and executing science observations which start when Sun's altitude is -9 degrees. In the end of the night, morning twilight flat images are taken, and when the slit of the dome is closed, calibration images are taken (bias and dark). We use a custom-made addition to \texttt{ACP}, which enables automatic restart (with no need for human intervention) of the observations after a period of safe time in case of a weather trigger (high wind and/or clouds).

Observing commands for \texttt{ACP} are contained in the text scripts, which are created automatically before the beginning of the night using the \texttt{SPOCK}\footnote{\url{www.github.com/educrot/SPOCK}} (SPeculoos Observatory sChedule maKer; \citealt{2020SPIE11445E..21S}). Typically, the telescope observes 1-2 targets per night (with an airmass upper limit of 2.4) and aiming at accumulating 100-200 hours for each SPECULOOS target.

To ensure continuous operations of SNO, the facility is regularly checked. Under the agreement with IAC, its staff performs a visual check of closure of the slit every day before sunrise. Every two weeks, IAC staff goes inside the facility to inspect for any signs of technical problems, which can not be seen remotely.

\subsection{Data flow}\label{subsec:data}

Depending on the exposure time (and thus on the number of the images), Artemis typically produces 4-16\,GB of data every night. After the end of the local night, science and calibration images (bias, dark and flat) are transferred to an archive at the University of Li\`ege. Later, from there the images are transferred to a storage server at the University of Cambridge (UK) where they are processed by SPECULOOS South data reduction pipeline \citep{2020MNRAS.495.2446M}. The same set of images is also transferred from the archive at the University of Li\`ege to a storage server at MIT where it is processed by the \texttt{Prose}\footnote{\url{www.github.com/lgrcia/prose}} pipeline \citep{2022MNRAS.509.4817G} to produce an independent set of light curves. 

Both pipelines perform standard image reduction steps (bias, dark, and flat-field corrections) and use an automated differential photometry iterative algorithm. Raw light curve of a target star is divided by an ‘artificial’ comparison light curve, which is constructed by weighting the sufficiently bright comparison stars according to their variability and distance to the target. Once the last night's data is processed, it is available at the SPECULOOS PORTAL \citep{2020SPIE11445E..21S}. Inspection of the light curves (including the ones from other SPECULOOS telescopes) at PORTAL is visual as we aim at detecting single transit events. Processed data is ready for an inspection before the beginning of the next night. Such a timely data processing allows us to spot any problems with the telescope and/or trigger follow-up observations of targets of interest.

When observations of a target from the SPECULOOS catalog are fully completed, a global light curve is created by applying the differential photometry algorithm to the entire time series at once (which can span several weeks or months). Global light curves are used to search for transit-like signals by the Box-fitting Least Squares method (BLS; \citealt{2002A&A...391..369K}) after removal of systematic effects and stellar variability in the data.

\section{Statistics of observations}\label{sec:weather}

\begin{figure}
    \centering
    \includegraphics[width=0.5\textwidth]{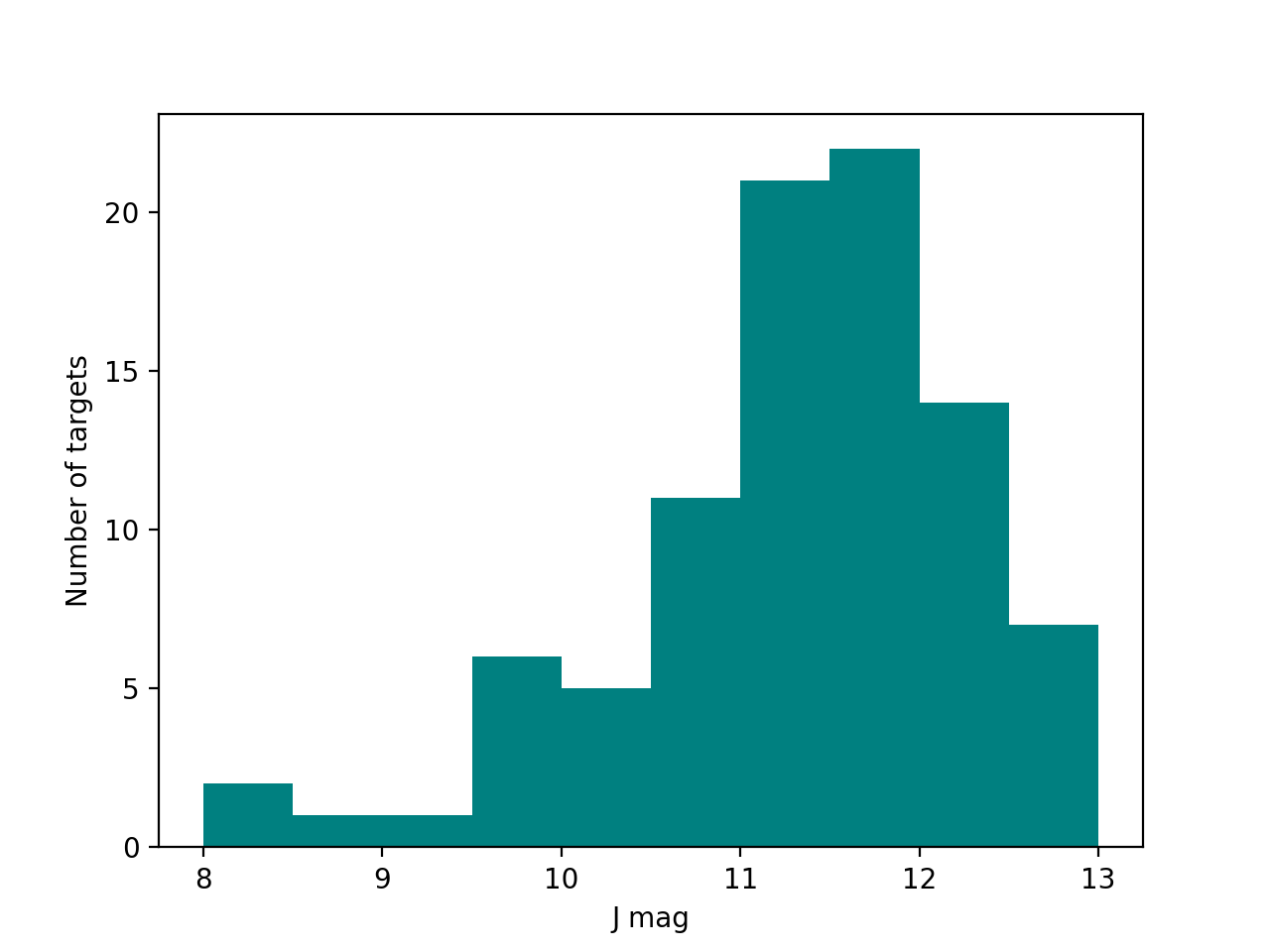}
    \includegraphics[width=0.5\textwidth]{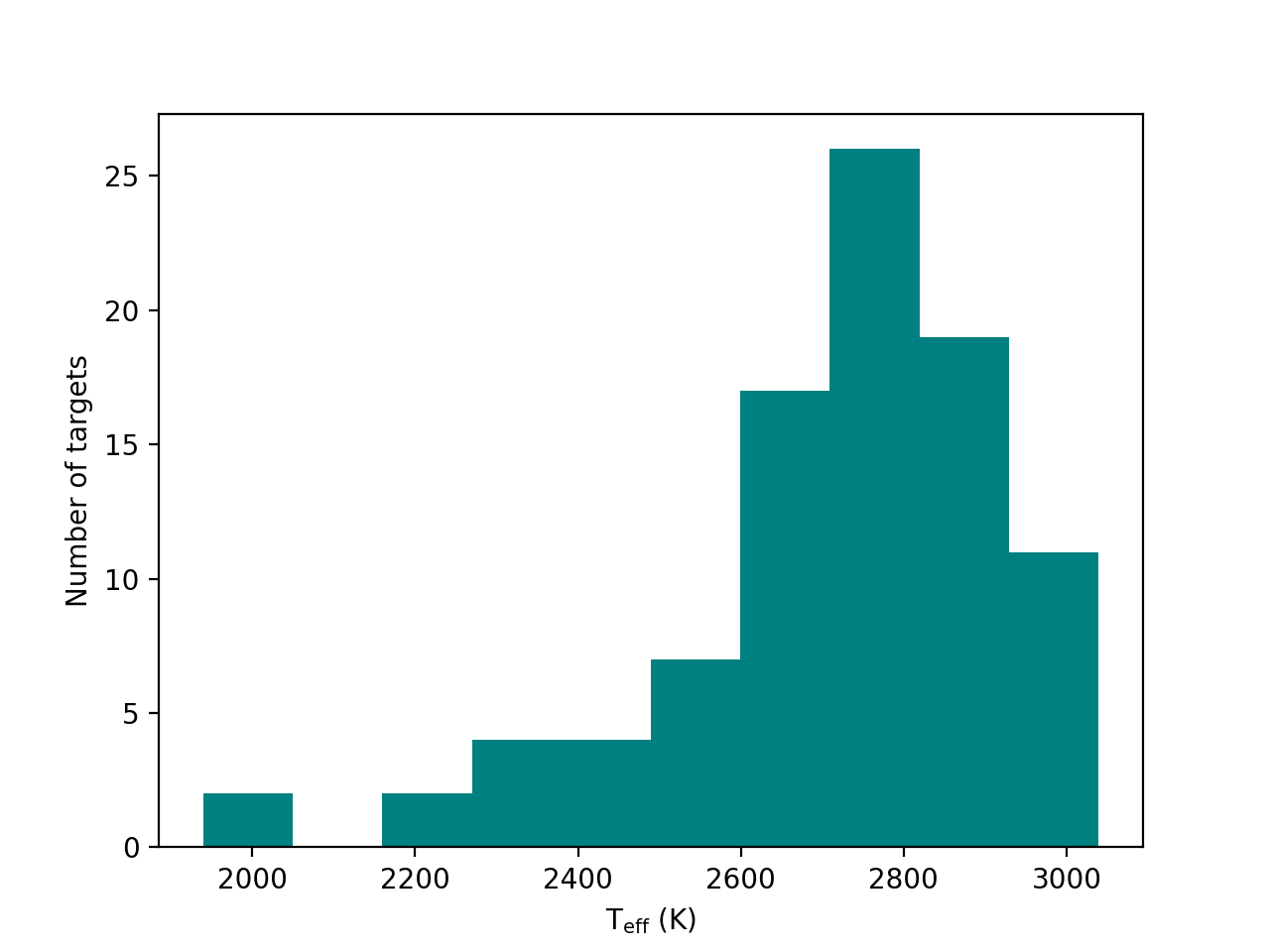}
    \caption{Top: the distribution of stellar magnitudes in $J$ band of all the targets, which were observed by the Artemis telescope from June 2019 to June 2022. Bottom: targets' distribution of effective temperatures.}
    \label{fig:target_dist}
\end{figure}

For a period of three years from June 2019 to June 2022, the Artemis telescope observed for 6000\,hr (on-sky time, not the total exposure time). Almost 90\% of all the targets are from SPECULOOS Program 1 (96 UCDs), whereas the remaining targets are split between Program 2, TESS follow-up, solar system small bodies follow-up observations, and educational programs. Program 1 features 365 UCDs, which are close-by and small enough for atmospheric studies of their possible planets with JWST. Program 2 focuses on UCDs later than M5 for which terrestrial planet detection should be within reach of TESS. The distribution of stellar magnitudes in $J$ band of all the observed targets is presented in Fig.~\ref{fig:target_dist}, along with the distribution of their effective temperatures. We refer the reader to the work of \citealt{2021A&A...645A.100S} for further details about the SPECULOOS target list and its programs. 

\begin{figure*}
    \centering
    \includegraphics[width=1.0\textwidth]{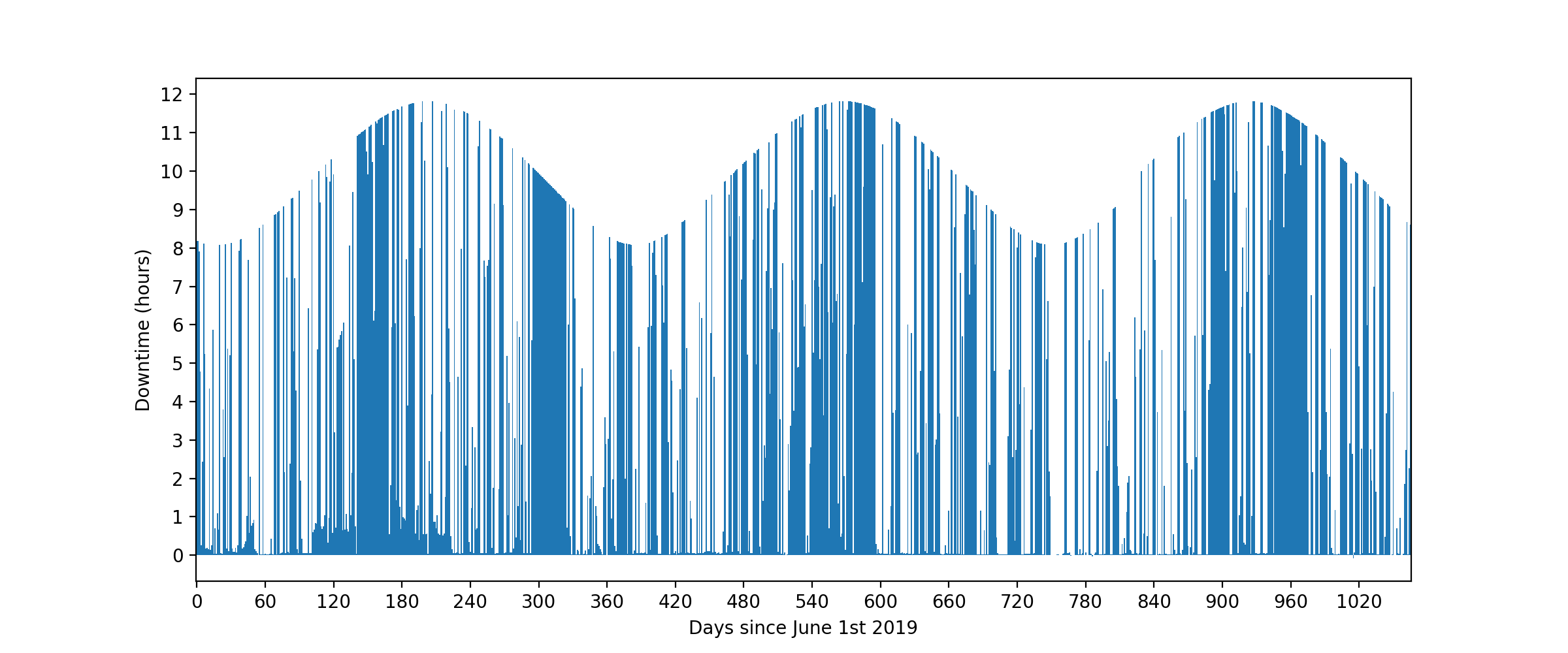}
    \caption{Artemis telescope downtime in hours starting from June 1st 2019. Maximum length of a bar depends on a duration of astronomical night reaching minimum in summer ($\sim$8\,hr) and maximum in winter ($\sim$12\,hr). Major blocks of downtime: technical interventions (block centered on day 150), observatory shutdown because of the COVID-19 pandemic (block centered on day 310), rain and snowstorms in winter seasons (blocks centered on days 580, 900, 960).}
    \label{fig:downtime}
\end{figure*}

Artemis' downtime during the first three-year period, defined as the ratio of the time on-sky to the sum of duration of all the nights, was 40\% (see Fig.~\ref{fig:downtime}). Percentage of nights with clement observing conditions using the Boltwood weather station measurements was 76\% (hence, the downtime according to the Boltwood station only was 24\%). These values are in agreement with the percentage of clear nights presented in Table~2 in \citealt{2021AJ....162...25A}, although higher than those recorded at IAC telescopes\footnote{http://research.iac.es/OOCC/about/statistics/} (17-18\% for the same period, which also includes downtime due to dust). We define observing night as clement if these conditions apply: there is no rain and cloud coverage (difference between temperature of the sky and ambient temperature is lower than -38\,$^{\circ}\mathrm{C}$), relative humidity is lower than 70\%, ambient temperature is 5\,$^{\circ}\mathrm{C}$ warmer than a dew point, wind speed is smaller than 45\,$\mathrm{km}\,\mathrm{hr^{-1}}$. If any weather factor reaches its threshold, \texttt{ACP} closes the dome automatically and reopens it if weather conditions are clement for at least 1\,hr. Difference in downtime between IAC telescopes and Artemis is due to more conservative weather thresholds employed at Artemis (in comparison with local decision made by human operators of IAC telescopes).

In contrast to the SPECULOOS telescopes in Chile and the SAINT-EX telescope in Mexico, SNO has an additional relatively small downtime because of the dust storms (\emph{calima} in Spanish). These events happen when strong seasonal winds carry sand and dust from the Sahara desert through the Canary Islands archipelago. At the altitude of the Teide Observatory, these events are sporadic and mainly occur in summer months. Such events display themselves as an increase of dust concentration in the air (which is \textit{not} necessarily accompanied by the strong wind). The \emph{calima} is strongly stratified with height in the lower atmosphere, with only $\sim$20\% of the total intrusions reaching the altitude of the observatory \citep{lak16}.

To monitor concentration of dust particles in the atmosphere, we have relied on a dust sensor\footnote{\url{www://stella-archive.aip.de/stella/status/status.php}} installed at the neighboring STELLA telescope \citep{2001AN....322..287S}. For calibration purposes we have also used the $\leq$10 micrometer particles (PM10) measurements carried out by the Spanish State Meteorological Agency (AEMet) on the neighbor atmospheric observatory, and managed by the IAC Sky Quality team\footnote{http://research.iac.es/OOCC/ciai-pm10}.  We also plan to install and use a commercial dust sensor Purple Air~PA-II\footnote{\url{www2.purpleair.com}} in June 2022, which will be placed on the roof of the Artemis control building. Our dust concentration threshold value from the STELLA sensor is 0.025 units, which correspond to a concentration of 150\,$\mu\mathrm{g/m^3}$ of calibrated PM10. We employ the threshold to prevent damage to the equipment, which can happen after accumulation of dust particles on the rails of the slit, inside the motors of the slit and dome, etc. During the three-year period of operations, we experienced 48 nights when concentration of dust was above the threshold for a least 30\,min. We have not operated the telescope on the nights when dust concentration was close or above the threshold at the beginning of the local night. This contributed to the overall downtime by increasing it from 24\% (Boltwood-only) to 30\%. In almost half of these cases, change of the dust level from clear to above 0.025 units happened in less than 10 hours.

As found by \citealt{1998NewAR..42..521J} and \citealt{1998NewAR..42..543F}, the behavior of Saharan dust is gray in the visible (450-870\,nm) and in the near-IR ($J$ and $H$ bands), meaning that moderately dusty nights should not pose a major obstacle for the differential photometry.

Besides 30\% weather-related downtime, additional 10\% time loss was associated with technical problems (see subsection~\ref{subsec:tech_downtime}). Our major blocks of downtime are caused by several factors (Fig.~\ref{fig:downtime}): 

\begin{itemize}
    \item Technical interventions by Astelco Systems during the first year after the installation of the telescope (e.g., a block centered on day 150 since June 1st 2019).
    \item Observatory shutdown because of the COVID-19 pandemic (a block centered on day 310).  
    \item Rain and snowstorms in winter seasons (blocks centered on days 580, 900, 960).
\end{itemize}

\subsection{Seeing}\label{subsec:seeing}

\begin{figure}[h]
\begin{center}
\includegraphics[width=0.5\textwidth]{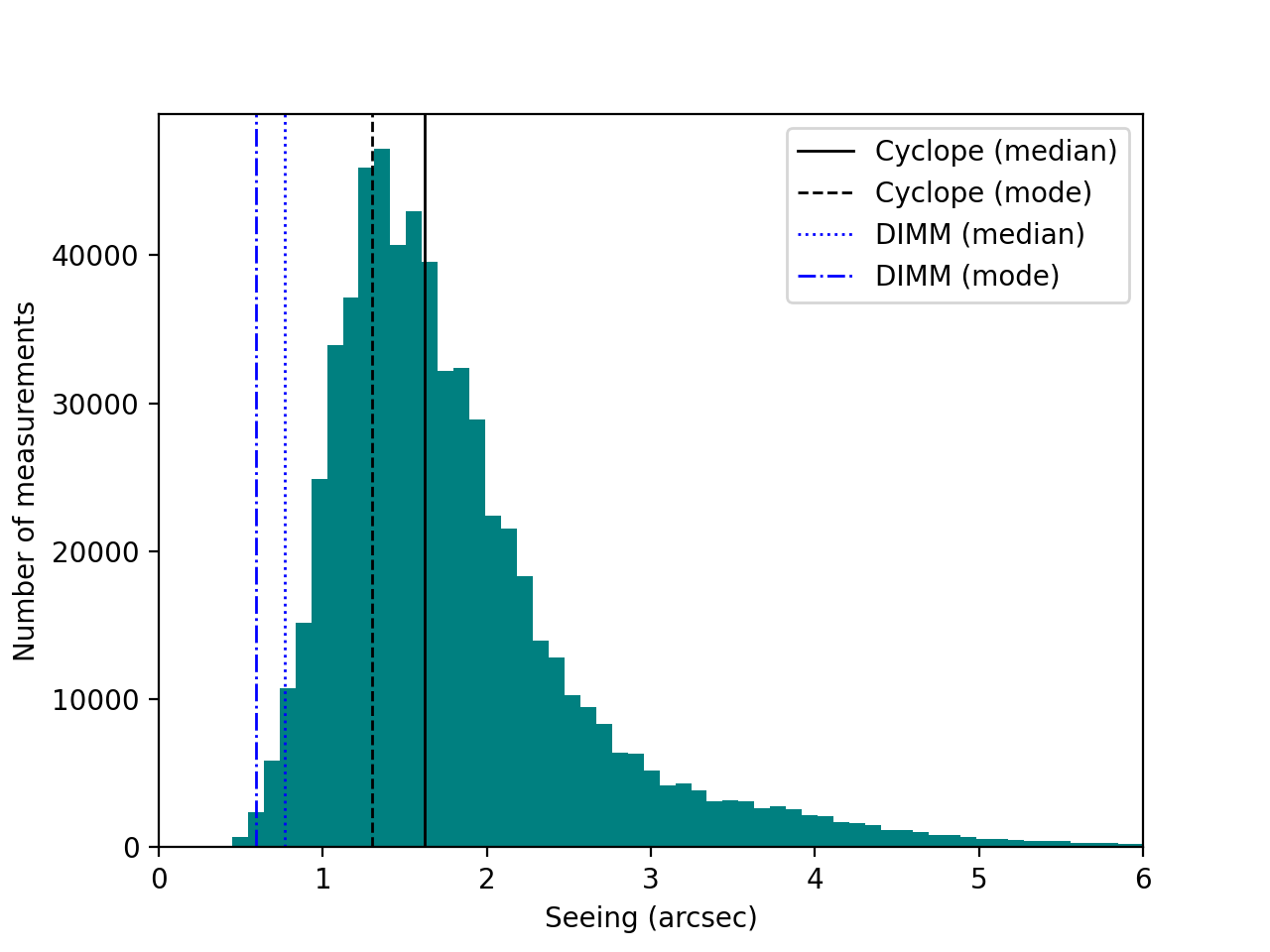}
\end{center}
\caption{The distribution of seeing measurements from April 2019 to April 2022 made by the Alcor Cyclope seeing monitor. The \textit{black} solid vertical line shows position of the median of the distribution (1.6\,arcsec), the \textit{black} dashed vertical line shows mode of the distribution (1.3\,arcsec). For comparison, values obtained with the IAC DIMM are presented (period February 2019 to January 2020; IAC Sky Quality Team, private communication). The \textit{blue} dotted vertical line shows position of the median of the distribution (0.77\,arcsec), the \textit{blue} dashed and dotted vertical line shows mode of the distribution (0.59\,arcsec). See Subsection~\ref{subsec:seeing} for more details.}
\label{fig:seeing}
\end{figure}

Regarding the seeing conditions, we rely on measurements\footnote{All measurements were obtained using Cyclope Software version 1.1.12 build 54} made with the Cyclope seeing monitor installed on the mast. The monitor is fixed and continuously observes the Polaris 50 times per sec, using a 1/125\,sec exposure time in a green filter. Then, a dedicated software measures the jitter of the Polaris and translates it to the line of sight and zenith seeing values. We found a good agreement between Cyclope and IAC Differential Image Motion Monitor (DIMM) seeing measurements made in April-June 2019 in tendency, behavior and log-normal statistical distribution, but with a bias of around 0.8\,arcsec in median values. According to our seeing observations from April 2019 until June 2022, the most frequent zenith seeing value (mode) was 1.3 arcsec, median value was 1.6 arcsec and its standard deviation was 0.9 arcsec (see Fig~\ref{fig:seeing}). Minimal registered seeing was
0.5\,arcsec. The same values obtained with the IAC DIMM (period February 2019 to January 2020; IAC Sky Quality Team, private communication) were 0.59\,arcsec (mode), 0.77\,arcsec (median), 0.48\,arcsec (standard deviation) and 0.19\,arcsec (minimal registered seeing), which are in agreement with equivalent values obtained with other techniques, such as Scintillation Detection and Ranging (SCIDAR; \citealt{gar11a, gar11b}). The discrepancies between Cyclope monitor and DIMM measurements may come from different factors, including a much larger contribution of the turbulence surface layer \citep{ver94} to the Cyclope monitor (due to its proximity to the ground). On the other hand, the 62\,degrees of zenith distance imposed by Polaris also imply that the turbulence models for zenith correction are edging (standard DIMMs at observatories usually measure up to 30\,degrees of zenith distance). In any case, for the purpose of this paper and the operation of SPECULOOS, the absolute seeing values are not as relevant as the seeing behavior. Further investigation on the precision and accuracy of the Cyclope seeing monitor will be carried out in a near future in collaboration with the IAC Sky Quality Team.

A median value of FWHM (full width at half maximum) of the stellar point spread functions (PSFs) measured by the SSO data reduction pipeline from the Artemis data over the course of three years, is 1.2\,arcsec. Discrepancy between median values from the Cyclope seeing monitor and from FWHM of stars comes from the fact that the seeing monitor records seeing whenever the Polaris is visible (which can happen when observing conditions do not allow the telescope to observe). Another factor is the difference in observing filters: the Cyclope monitor records seeing in a green filter (central wavelength 550\,nm), while the majority of observations with Artemis are conducted in $I+z'$ filter (central wavelength $\sim900$\,nm).

\subsection{Major technical problems}\label{subsec:tech_downtime}

Though a periodic telescope maintenance decreases a total number of technical problems, sporadic major issues occur. The Artemis telescope suffered from a problem with an excessive optical aberration (spherical), which was fixed by Astelo Systems in the end of 2019. Failure of a thermo-electrical cooling component of the Andor CCD camera forced us to use a replacement CCD camera for a period of 6 months in 2020 (until the science camera was repaired and installed back at the telescope). 

Because of harsh winter season conditions at the Teide Observatory, control room and dome areas experienced minor water leakages during the first winter of operations (2019-2020). Problematic areas were sealed and water-proofed. Other weather-related incidents were caused by ice formed on the slit of the dome (which prevented the slit opening) and inside the slit rails (which prevented closure of the slit in April 2022). Ice formation and frozen precipitation can happen from October to May, with maximum probability in March \citep{cas18}.

\section{Photometric performance}\label{sec:photometry}

All SPECULOOS telescopes aim at detecting single transits from terrestrial planets around UCDs. Depending on the mutual sizes of planets and host stars, transit depths might differ from a few 0.1\% up to several per cent. Reaching this goal requires photometric precision of $\sim$0.1\% (1\,mmag), which is typically obtained with the Artemis telescope. As an illustration of this precision, we provide light curves of a known transiting exoplanet around a UCD (TRAPPIST-1) and some of the targets from the SPECULOOS input catalog.

\begin{figure}[h]
    \centering
    \includegraphics[width=0.475\textwidth]{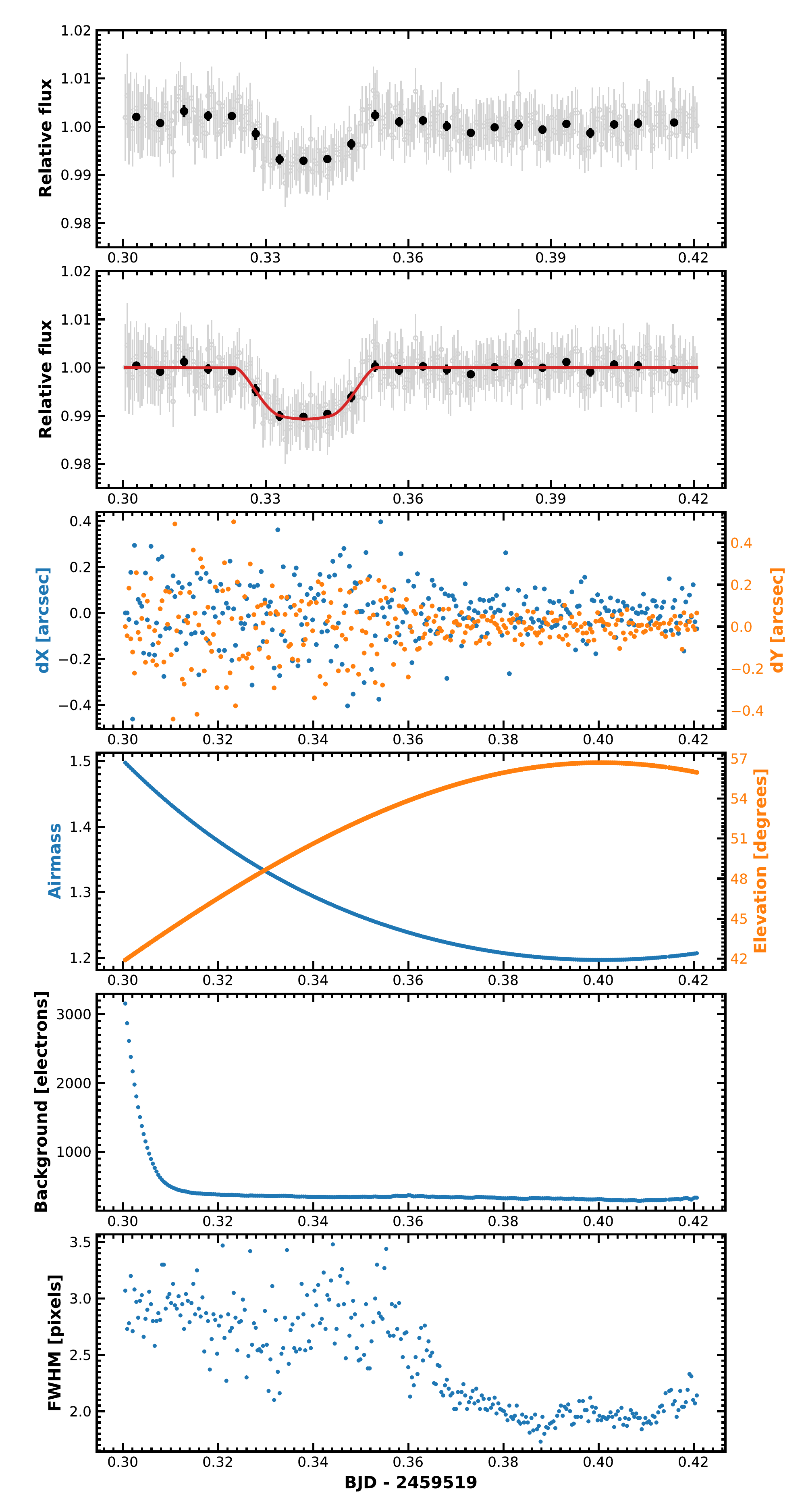}
\caption{Differential light curve of TRAPPIST-1\,b (top panel: before detrendeing, second to top panel: after detrending. The solid red line represents the best-fit model). The data was obtained with the Artemis telescope on 31 Oct 2021 in $I+z'$ filter with 23 sec exposure. Evolution of systematics throughout the observing run is presented in the remaining panels: shift of stars' positions on the CCD along X and Y axes (dx, dy), change of airmass and elevation, evolution of sky background and FWHM (full width at half maximum of the stellar point spread function).}
    \label{fig:trappist-1h_LC}
\end{figure}

TRAPPIST-1\,b is the shortest-period planet in the TRAPPIST-1 system ($J$=11.4\,mag, $I$=14.0\,mag) with a radius of $1.12\pm0.014\,\mathrm{R_{E}}$, where $\mathrm{R_{E}}$ is the radius of Earth \citep{2021PSJ.....2....1A}. Its transit was observed on 31 Oct 2021 in $I+z'$ filter with 23 sec exposure. Its light curve and evolution of systematics throughout the observing run (shift of stars' positions on the CCD along X and Y axes, change of airmass, sky background and FWHM of PSFs) are presented in Fig.~\ref{fig:trappist-1h_LC}. The transit is clearly detectable and its significance is 12-$\sigma$. The light curve was detrended (linear trend removal) in order to minimize the RMS of the best-fit residuals, which are 620\,ppm per 7.2\, min bins. Note sub-pixel precision of the tracking coupled with the DONUTS software guiding algorithm.

\begin{figure}[h]
    \centering
    \includegraphics[width=0.475\textwidth]{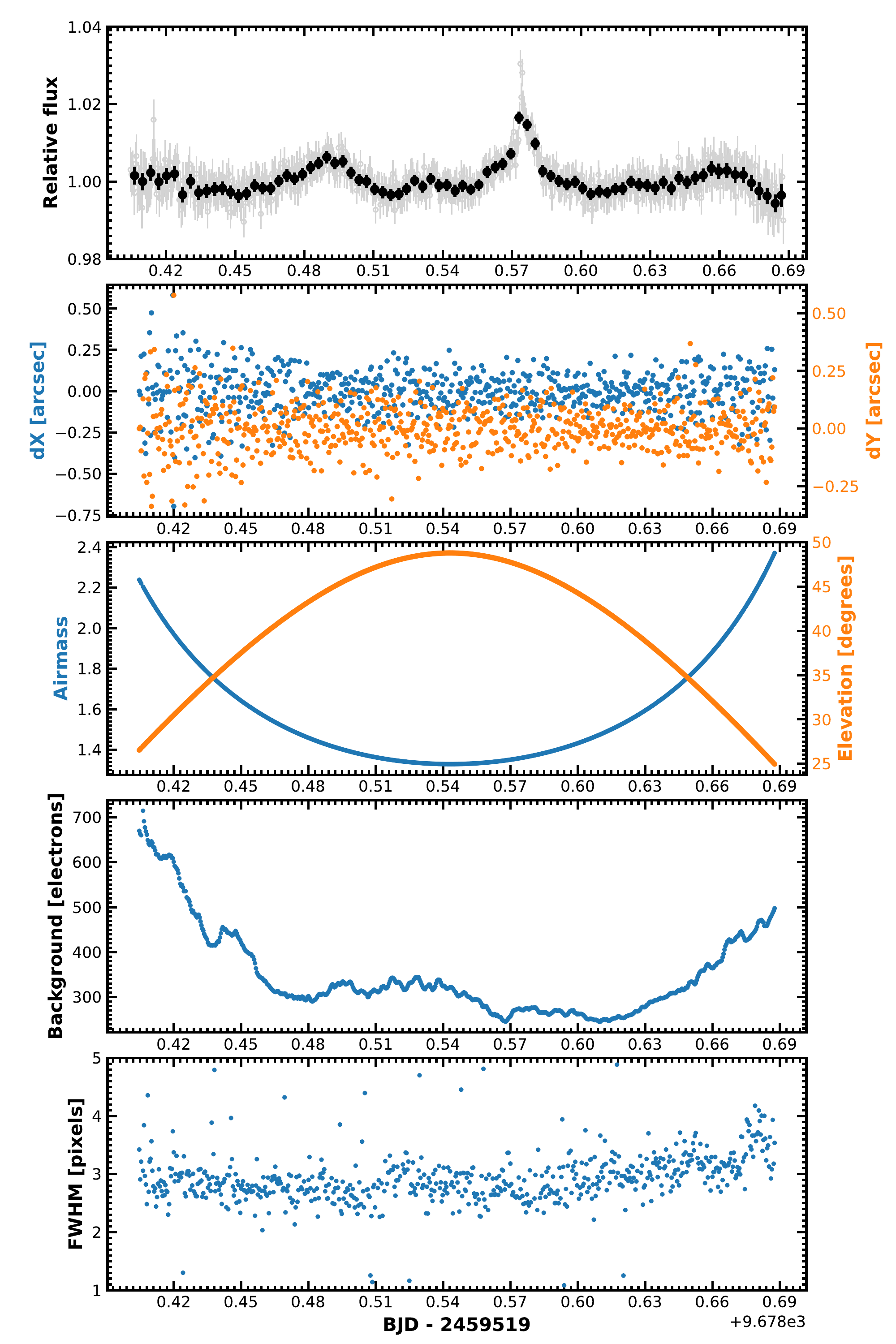}
    \caption{Light curve of Sp1256-1257 (VHS 1256-1257; $J$=11.0\,mag, $I$=14.3\,mag) obtained with the Artemis telescope on 8 April 2022. Sp1256-1257 is a triple brown dwarf system \citep{2015ApJ...804...96G,2016ApJ...818L..12S}.}
    \label{fig:var_flare}
\end{figure}

Sp1256-1257 (VHS 1256-1257; $J$=11.0\,mag, $I$=14.3 mag) is a triple brown dwarf system \citep{2015ApJ...804...96G,2016ApJ...818L..12S}. The primary of this system is an equal-magnitude binary which components spectroscopically determined to be M7.5$\pm$0.5. It shows complex photometric variability, including flaring activity. One of its light curves is presented in Fig.~\ref{fig:var_flare}, where a 3\,\% flare is visible. Such type of light curves makes possible precise determination of periodic features (e.g., rotational periods of UCDs), and the exploration of the possible relationship between flaring activity and rotation period (e.g, see \citealt{2022MNRAS.tmp.1043M}).

\begin{figure}[h]
    \centering
    \includegraphics[width=0.5\textwidth]{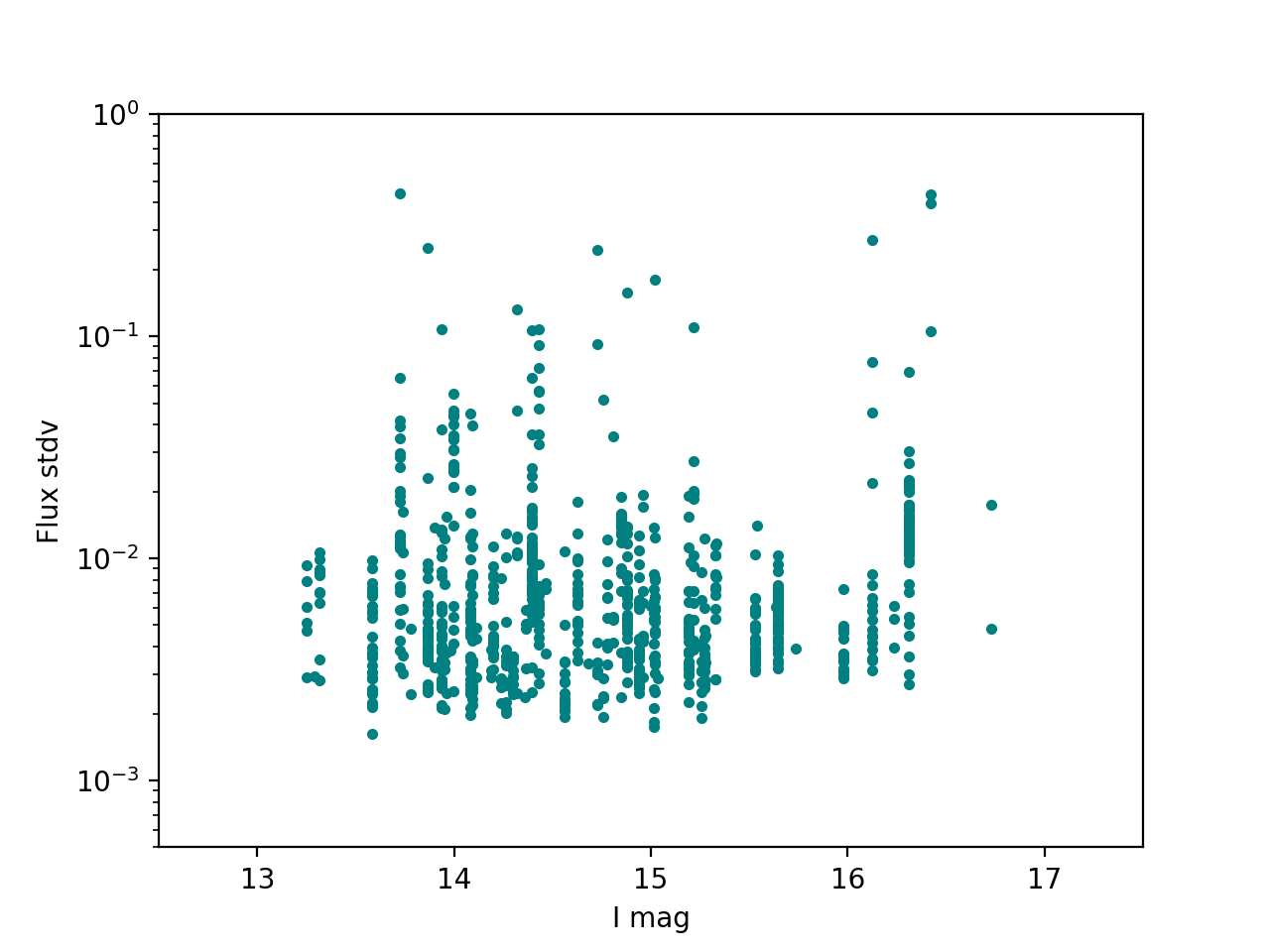}
    \caption{Flux standard deviation of all UCDs' lightcurves observed with the Artemis telescope in $I + z'$ filter from June 2019 to June 2022. The vertical lines correspond to different flux standard deviations on different nights, but for the same targets.}
    \label{fig:flux_stdv}
\end{figure}

We also analyzed individual light curves processed by the SSO data reduction pipeline of every SPECULOOS target, which was observed with the Artemis telescope in $I+z'$ band for the last three years. For every light curve from each night spanning more than 120\,min, we calculated flux standard deviation after the differential photometry and normalization (with no data binning, sigma-clipping and/or trend removals). A global set of flux standard deviations as a function of target stellar magnitude in Cousins $I$ band are presented in Fig.~\ref{fig:flux_stdv}. The vertical lines on the plot correspond to different flux standard deviations on different nights, but of the same target. The vertical scatter of these points is explained by either the intrinsic variability of the target (e.g, flares, rotational modulations), or by variations in the observing conditions from night to night. Our typical photometric precision is $0.5\%$ (median of a set of flux standard deviations of the observed targets). For bright non-variable targets, we are able to reach flux standard deviations $0.2\%$ with a typical exposure time of 25\,sec.

We note that second-order extinction effects due to highly variable absorption by atmospheric water vapor can cause changes of differential flux of very red targets mimicking or hiding transits and affecting the long-term photometric variability studies of our targets (\citealt{2020MNRAS.495.2446M}, Pedersen et al. 2022, submitted). This problem is mitigated at SSO by using precise high-cadence (every 2 minutes) precipitable water vapor (PWV) measurements from a microwave radiometer optimized for measuring PWV in dry conditions (from 0\,mm to a saturation value of 20\,mm, within an accuracy of 0.1\,mm). Such PWV measurements are used to correct differential light curves from SSO as part of the automatic pipeline.

Lower-cadence (every 30 minutes) PWV measurements with a precision of 1\,mm are available from a GPS-based system located at the Teide Observatory \citep{2016SPIE.9910E..0PC}. Much higher cadence (every 1.5 minutes) and more precise PWV values will be also available from mid-2022 after the testing of a newly installed radiometer by the Japanese company Furuno\footnote{\url{www.furuno.com}} at the Teide Observatory. Currently, data from the Artemis telescope does not undergo any PWV correction. However, we plan to compare PWV correction based on the data from GPS and from the Furuno radiometer in the future and correct differential light curves as part of the automatic pipeline.

While most of the observing time is dedicated to targets from the SPECULOOS catalog, Artemis does follow-up observations of other noteworthy exoplanets, including the ones from TESS and Kepler/K2 \citep{2014PASP..126..398H}.  We refer the reader to the work by \cite{2022MNRAS.514.4120G} featuring Artemis' light curve of a sub-Neptune orbiting a mid-M dwarf TOI-2136 and to the work by \cite{2020AJ....160..172N} showing a transit of an Earth-sized planet around an M3.5 dwarf observed with Artemis and several SSO telescopes. 
Artemis also does follow-up observations of solar system minor bodies, e.g., photometric monitoring of asteroids, which display cometary activity \citep{2021MNRAS.505..245D} and observations of occultation of stars by asteroids \citep{2021DPS....5350305F}.

\section{Complementary science}\label{sec:complementary}

Upon completion of the survey, SPECULOOS telescopes will have observed almost 70\,$\mathrm{deg^2}$ of sky for 100-200\,hr with remarkable photometric performance and high cadence. In order to fully utilize the potential of this data set, we initiated an automatic detection of "moving objects", i.e., objects which cross a field of view of the telescope. Such moving objects are typically asteroids and comets (small bodies), which are pristine remnants of the solar system formation.

To find these objects, we employ digital tracking (also known as synthetic tracking, or shift and stack method; \citealt{1995ApJ...455..342C,2014ApJ...782....1S,2014ApJ...792...60Z,2015AJ....150..125H}). This approach relies on shifting and stacking of individual astronomical images according to a motion vector of a moving body in order to increase the object’s signal in the data. This approach allows the detection of smaller and/or more distant objects in the solar system compared to a conventional technique, which determines if an object exhibits consistent movement from one image to the next (and where the moving object’s signal-to-noise ratio is limited to that of a single exposure). Our dedicated GPU-accelerated (Graphics Processing Unit) digital tracking pipeline is based on the \texttt{Tycho Tracker} software\footnote{\url{www.tycho-tracker.com}}. The pipeline searches for moving objects in the archival images and in the last night's data (to be able to trigger follow-up of noteworthy objects during the next observing night). The pipeline is able to confidently detect moving objects as faint as $V\sim$23.0\,mag. The first results of the pipeline application on a set of archival SPECULOOS fields will be presented in Burdanov et al. 2022, submitted.


\section{Discussion and conclusions}\label{sec:discuss_conclusions}

We have presented Artemis: the first telescope of the SPECULOOS Northern Observatory (SNO), and its development over the course of the first three years of operations at the Teide Observatory on the island of Tenerife. According to our weather station measurements, percentage of nights with clement observing conditions was 76\%. However, our actual downtime was 40\% because of additional time loss associated with technical problems, nights lost due to dust storms and observatory shutdown because of the COVID-19 pandemic. We plan to use our dust sensor and implement automatic slit closure and re-opening using its data. Thanks to this and fewer major technical problems, we expect the downtime to decrease during the next years.

The Artemis telescope demonstrates remarkable photometric precision, allowing it to be ready to fulfill its main goal -- finding new transiting terrestrial exoplanets around UCDs. Over the period of the first three years after the installation, we observed 96 objects from the SPECULOOS target list for 6000\,hours with a typical photometric precision of $0.5\%$, and reaching a precision of $0.2\%$ for relatively bright non-variable targets with a typical exposure time of 25\,sec. We plan to compare PWV correction based on the data from GPS and from the Furuno radiometer in the future and correct differential light curves as part of the automatic pipeline. We expect further improvements of our photometric precision after applying the PWV correction. Though no new planets transiting UCDs were confirmed yet, the SPECULOOS survey continues as currently less than 10\% of targets were fully observed.



\acknowledgments

\textit{Acknowledgments:}
J.d.W. and MIT gratefully acknowledge financial support from the Heising-Simons Foundation, Dr. and Mrs. Colin Masson and Dr. Peter A. Gilman for Artemis, the first telescope of the SPECULOOS network situated in Tenerife, Spain. The ULiege's contribution to SPECULOOS has received funding from the European Research Council under the European Union's Seventh Framework Programme (FP/2007-2013) (grant Agreement n$^\circ$ 336480/SPECULOOS), from the Balzan Prize and Francqui Foundations, from the Belgian Scientific Research Foundation (F.R.S.-FNRS; grant n$^\circ$ T.0109.20), from the University of Liege, and from the ARC grant for Concerted Research Actions financed by the Wallonia-Brussels Federation. 

Authors would like to thank the anonymous reviewer for their time and attention. The constructive comments we received, helped us to improve the quality of the paper.

The SPECULOOS North consortium would like to thank IAC telescope operators (Técnico de Operaciones Telescópicas), General and Instrumental maintenance teams for their support on site, IAC Sky Quality Team for providing useful comments and access to PWV measurements, THEMIS solar telescope team and Dr.~Carlos Dominguez for their invaluable help during the installation of the Artemis telescope.

%

\facilities{SPECULOOS Northern Observatory, SPECULOOS Southern Observatory, Teide Observatory}


\software{\texttt{DONUTS} \citep{2013PASP..125..548M}, \texttt{ACP} (\url{www.acp.dc3.com}), \texttt{SPOCK} (\url{www.github.com/educrot/SPOCK}), \texttt{Prose} (\url{www.github.com/lgrcia/prose}), \texttt{Tycho Tracker} (\url{www.tycho-tracker.com})}

\bibliography{Burdanov}



\end{document}